\documentclass[a4paper]{aa}
\usepackage{graphics,txfonts}

\def\teff{$T_{\rm eff}$}
\def\lgg{$\log g$}

\def\vs{$v_{\rm e}\sin i$}

\newcommand{\bs}{$\langle B \rangle$}
\newcommand{\kms}{km\,s$^{-1}$}

\newcommand{\li}{\ion{Li}{i}}
\newcommand{\hd}{HD\,116114}
\newcommand{\sm}{{\tt SYNTHMAG}}
\newcommand{\mt}{$^2S$--\,$^2P$}

\newcommand{\figps}[1]{\resizebox{\hsize}{!}{\rotatebox{0}{\includegraphics{#1}}}}

\newcommand{\fifps}[2]{\centering\resizebox{#1}{!}{\includegraphics{#2}}}

\newcommand{\beq}{\begin{equation}}
\newcommand{\eeq}{\end{equation}}

\begin{document}

\title{The Paschen-Back effect in the Li~{\sc i} 6708~\AA\ line and
the \\ presence of lithium in cool magnetic Ap stars%
\thanks{Based on observations collected at the European Southern Observatory,
Paranal, Chile (ESO program 68.D-0254 and programs 072.D-0138, 077.D-0150 retrieved 
through the ESO Archive)}}

\author{O. Kochukhov}
 
\offprints{O. Kochukhov, \email{oleg@astro.uu.se}}

\institute{Department of Astronomy and Space Physics, Uppsala University, SE-751 20, Uppsala, Sweden}

\date{Received / Accepted }

\abstract%
{
A number of cool magnetic Ap stars show a prominent feature at $\lambda$~6708~\AA.
Its identification with \li\ remains controversial due to a poor
knowledge of the spectra of rare-earth elements that are strongly
enhanced in peculiar stars and can potentially provide an alternative
identification.
}
{
We suggest to investigate the 6708~\AA\ line in Ap stars with strong
magnetic fields. In these objects the magnetic broadening and splitting provides an
additional powerful criterium for line identification, allowing to use the whole
line profile instead of a mere coincidence of the observed and predicted wavelength.
}
{
Due to a small separation of the \li\ doublet components, their magnetic 
splitting pattern deviates from the one expected for the Zeeman effect even in
relatively weak fields. We carry out detailed calculations of the transition
between the Zeeman and Paschen-Back regimes in the magnetic splitting of the \li\ line
and compute polarized synthetic spectra for the range of field strength expected
in Ap stars. Theoretical spectral synthesis is compared with the high-resolution
observations of cool Ap stars HD\,116114, HD\,166473 and HD\,154708, which have a mean field 
strength of 6.4, 8.6 and 24.5~kG, respectively, and show a strong 6708~\AA\ line.
}
{ 
High-resolution spectra for the 6708~\AA\ region are analysed for 17 magnetic
Ap stars. The presence of the 6708~\AA\ line is confirmed in 9 stars and
reported for the first time in 6 stars. The strength of the \li\ doublet does
not correlate with absorption features of any other element. The stars HD\,75445 and
HD\,201601 provide an extreme example of the two objects which are dissimilar
with respect to the 6708~\AA\ line but very close in the atmospheric parameters and
abundances of other elements. We
demonstrate that the observed profiles of the 6708~\AA\ line in the strong
field stars HD\,116114, HD\,166473 and HD\,154708 correspond rather well to the
theoretical calculations assuming the \li\ identification. Inclusion of the
Paschen-Back effect improves the agreement with observations, especially for
HD\,154708.
}
{
Results of our study confirm the \li\ identification proposed for the 
6708~\AA\ line in cool Ap stars.
}

\maketitle

\section{Introduction}
\label{intro}

Chemically peculiar A and B stars (also know as Ap/Bp stars) represent unique
astrophysical laboratories for investigation of the hydrodynamic processes,
rotation and magnetism in the upper main sequence stars. Cool magnetic Ap stars
(\teff\,$\la$\,8000~K) are especially interesting due to their slow rotation,
the presence of non-radial oscillations, strong global magnetic fields, and
remarkable richness of the optical spectra. Abnormal strengths of spectral
lines  in these stars reveal large deviations of the stellar surface chemistry
from the scaled solar abundance pattern. Modern abundance analyses (see a summary
by Ryabchikova \cite{R05}) typically find that the light elements are
underabundant, the iron-peak elements are mildly enhanced or close to solar,
whereas huge overabundances are reported for the heavy elements, especially
rare-earths. Adding complexity to this observational picture, chemical elements
are distributed inhomogeneously in many Ap stars, showing both a non-uniform
horizontal structure (chemical spots) and different distribution with height
(stratification) in the stellar atmosphere. Combined with a prominent Zeeman
broadening and splitting of spectral lines, these unusual properties make the
appearance of Ap-star spectra dramatically different compared to the normal stars
of similar spectral classes. In particular, the line density is greatly
increased, with the spectral regions accessible to typical modern CCD
spectroscopy populated by thousands of unidentified lines of the rare-earth
elements (REEs). This obviously creates significant problems for identification
and measurement of the lines of astrophysically interesting ions. 

The history of identification and analysis of lithium in Ap stars is
characteristic of the difficulties encountered in the studies of elements with a
small number of absorption lines. The feature at the position of the resonance
\li\ doublet at $\lambda$ 6707.76 and 6707.91~\AA\ (hereafter referred to as
\li\ 6708~\AA) was originally reported for several Ap stars based on
low-resolution, poor $S/N$ observations (e.g., Wallerstein \& Merchant
\cite{WM65}). Modern observations with electronic detectors (Faraggiana et al.
\cite{FGD96}; Polosukhina et al. \cite{polo}) suggested the presence of the
resonance Li line in about a dozen stars, and also uncovered a conspicuous
variation of this line with the rotational phase in a few objects. The lithium
abundance exceeding the solar concentration of this element by 2--3 orders of
magnitude is required to reproduce the observed equivalent width of the
6708~\AA\ line (Shavrina et al. \cite{shav}; Polosukhina \& Shavrina \cite{PS07}). 
Moreover, up to 6~dex lithium
overabundance in the areas around magnetic poles is deduced in the
spatially-resolved Doppler imaging investigation of the Li spots in HD\,83368
(Kochukhov et al. \cite{KDP04}) and in HD\,3980 (Drake et al. \cite{hd3980}).  
Such an unusual behaviour, combining an
extreme lateral inhomogeneity and a large overabundance, is often found for the
REEs in Ap stars, but is uncharacteristic of the light elements. Consequently, the
\li\ identification of the 6708~\AA\ feature has been often questioned  (e.g.
Nesvacil et al. \cite{NHM04}), citing incompleteness of the currently available
REE line lists. Indeed, an element identification hinging on a single line can
be incorrect and resulting abundance can be significantly overestimated if an
unknown REE line overlaps by chance  with the expected position of the resonance
lithium line. Interestingly, recent extension of the \ion{Ce}{ii} line list has
solved a similar ``Li problem'' for the s-process enriched low-mass post-AGB
stars, where the absorption previously attributed to the redshifted \li\ line
has turned out to be due to \ion{Ce}{ii} 6708.099~\AA\ (Reyniers et al.
\cite{rey}).

Inconclusive situation with the identification of Li in cool Ap stars warrants
a new look at this problem. The previous studies were largely concerned with
measuring Li abundance and interpreting the line strength variations, but have
not fully explored the line profile information available in the
high-resolution observations of the slowly rotating Ap stars. Here we propose that
an additional interesting constraint and, possibly, an ultimate confirmation of
the Li identification, can be obtained through the study of the 6708~\AA\ line 
shape in stars with strong magnetic field. As pointed out by Mathys \& Lanz
(\cite{ML95}), line identification in the Ap stars with resolved Zeeman split lines
can be strengthened by comparison of the observed and expected magnetic
splitting patterns. Effectively, in stars with strong fields not only the
central line position but also the whole line profile shape, determined by the
number and relative strengths of the $\pi$ and $\sigma$ components, can serve as 
a consistency check in the cases where line identification is doubtful. The first
practical application of this strategy was presented by Ryabchikova et al.
(\cite{RRKB06}). These authors used observations of the resolved Zeeman-split lines
in the strongly magnetic Ap star HD\,144897 for a major reclassification of the
\ion{Nd}{iii} spectrum, thereby substantially increasing the number of
\ion{Nd}{iii} lines with precise wavelengths and atomic parameters. 

The aim of the present paper is to study the formation of the \li\ resonance doublet in
the slowly rotating cool Ap stars taking into account its special behaviour in a strong magnetic field. 
It is well known that for the kG-strength magnetic fields typically found in Ap
stars, the splitting of the Li line exhibits substantial deviation from the normal
linear separation of the Zeeman components due to the partial Paschen-Back effect (e.g. Maltby
\cite{M71}; Mathys \cite{M91}). The importance of the Paschen-Back effect for the 
profiles of individual multiplets of
iron-peak elements was demonstrated by Mathys (\cite{M90}), Landolfi et al.
(\cite{LBLL01}) and Stift \& Leone (presentation at the Vienna CP\#Ap Workshop, 
September 2007), however no study has previously 
incorporated the Paschen-Back splitting in the \li\ line formation calculations for
magnetic A-type stars. 

In Sect.~\ref{pbe} we address the problem
of the departure of magnetic splitting in the \li\ 6708~\AA\ line from the simple Zeeman
pattern. We present calculation of the partial Paschen-Back effect in the Li doublet
and demonstrate how transition from the weak-field Zeeman to the strong-field
Paschen-Back regimes manifests itself in the theoretical profiles of the Li line
computed for the conditions typical of the cool Ap-star atmospheres.
Sect.~\ref{lihd} gives an overview of the available observations of the Li line in Ap
stars with strong magnetic fields and presents a spectral synthesis modelling of the
6708~\AA\ feature in Ap stars HD\,116114, HD\,166473 and HD\,154708, which turn out to be
the most suitable objects for the purpose of our investigation. Finally,
conclusions of the study are summarised in Sect.~\ref{conc}. 

\section{The Paschen-Back effect for the \mt\ transition}
\label{pbe}

In the usual situation of the Zeeman effect, the splitting produced by the
magnetic field is much smaller compared to the energy separation between
different $J$-levels of a multiplet. In this case the influence of the magnetic
field can be considered as a small perturbation to the fine-structure
Hamiltonian describing the system in the absence of the field. However, when
the magnetic splitting becomes comparable to the fine-structure separation, the
perturbation theory of the Zeeman effect is no longer valid. In this,
so-called Paschen-Back regime, the Hamiltonians corresponding to the
spin-orbit interaction and to the magnetic field effects should be diagonalized
simultaneously. A comprehensive derivation of the theory of the Paschen-Back
effect in atomic systems is given by Landi Degl'Innocenti \& Landolfi
(\cite{LL04}, \S\ 3.4) and will not be repeated here. It suffices to say
that the expressions for the splitting and strengths of the $\pi$ and $\sigma$
components can be derived under the assumption of $L$-$S$ coupling, but
evaluation of these expressions has to be performed numerically for all but the
simplest terms.

The \li\ doublet is formed by the transition between the $^2S$ and $^2P$ levels.
The upper level exhibits the fine-structure splitting into the levels with
$J=1/2$ and $3/2$, separated by $\Delta E_J=0.337$~cm$^{-1}$. The behaviour of the energy
levels as a function of the magnetic field strength can be characterized by the
quantity
\beq
\omega=\frac{\displaystyle ehB}{\displaystyle 4\pi m_{\rm e}c\beta},
\eeq 
where $B$ is the magnetic field strength, $\beta$ is the multiplet separation
constant ($\beta=\Delta E_J/3=0.112$~cm$^{-1}$ for the $^2P$ term in \li) and 
other constants have their usual meaning. In the regime of the Zeeman effect
($\omega$\,$\ll$\,1), the magnetic splitting increases linearly with the field
strength. For $\omega$\,$\sim$\,1, which for the \li\ 6708~\AA\ doublet corresponds 
to $B$\,$\sim$\,1~kG, the eigenvalues start to perturb each other, and linearity of the
splitting pattern is lost. This regime is often called the incomplete or partial
Paschen-Back effect. Finally, for $\omega$\,$\gg$\,1 (the complete Paschen-Back effect)
the spin-orbit interaction can be treated as a small perturbation in comparison
with the magnetic splitting. For the \li\ 6708~\AA\ line the complete
Paschen-Back triplet splitting pattern will emerge for $B$\,$\ga$\,12~kG. Each
component of this normal Zeeman pattern is split according to the multiplicity
of the electronic levels.

\begin{figure}[!th]
\figps{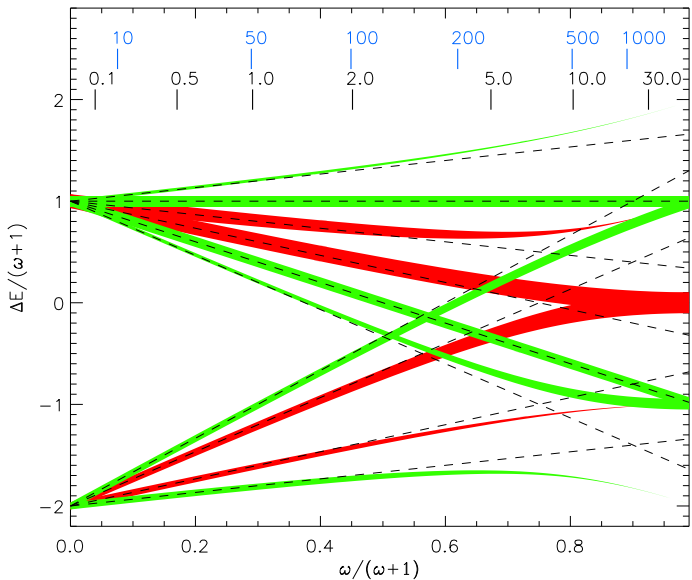}
\caption{The strength and separation of the magnetic components of the 
\mt\ multiplet as a function of the magnetic field strength. The
relative energy $\Delta E$ (vertical axis) is given in units of the multiplet 
splitting constant $\beta$. Both axes are divided by $\omega+1$. Dashed lines
show splitting for the linear Zeeman effect. Shaded curves correspond to the
magnetic splitting calculated taking into account the Paschen-Back effect. The
width of the curves is proportional to the strength of the respective $\pi$
(darker curves) and $\sigma$ (lighter curves) components. The vertical bars show
the magnetic field strength (in kG) corresponding to the fine-structure splitting 
in the Na~D lines (upper row) and the \li\ resonance doublet (lower row).
}
\label{fig1}
\end{figure}

These considerations show that substantial deviations from the linear magnetic
splitting characteristic of the simple Zeeman effect are expected to occur in the
\li\ 6708~\AA\ line for the magnetic field strengths of a few kG. Such fields are
common at the surfaces of magnetic Ap stars and in the magnetic regions on the Sun
and active late-type stars. Thus, an investigation of the \li\ resonance doublet
requires detailed Paschen-Back treatment of the magnetic splitting for essentially
all types of magnetic stars. However, since the experimental work of Paschen and Back
(Back \cite{B12}), little attention has been paid in the context of astrophysical
studies to the unusual response of the Li line to a strong magnetic field. Based on the
quantum mechanical calculations, Darwin (\cite{D27}) and Darwin (\cite{D28})
presented rules for forming chains of non-linear equations from which the parameters of
the line splitting in the partial Paschen-Back regime could be derived. In a series of
papers, published between 1930 and 1941 (Green \cite{G30,G41}), Green and
collaborators applied the Darwin's formalism to the interpretation of the laboratory
measurements of the emission spectra of various ions in strong magnetic fields. The
third paper in the series (Green \& Loring \cite{GL36}) examined the strong field
behaviour of the \mt\ multiplet of \li. Using the results of Darwin's calculations, Traub
(\cite{T68}), Engvold et al. (\cite{EKM70}) and  Maltby (\cite{M71}) 
studied formation of the resonance \li\ line in
sunspots. They have carried out the line profile computations for the \li\ doublet under
the assumption of the Milne-Eddington atmosphere. 
Mathys (\cite{M91}) presented another theoretical calculation of the Paschen-Back
splitting in the lithium line. Mathys (\cite{M90}) and Nesvacil et al. (\cite{NHM04})
stressed the importance of taking this effect into account in the analyses of the Li feature
in Ap stars. However, several recent studies of the \li\ line
in magnetic stars, even those supposedly presenting refined modelling of the Li 
transition in the magnetic fields exceeding 3--5~kG (e.g. Polosukhina \& Shavrina
\cite{PS07}; Leone \cite{L07}), have failed to include the Paschen-Back effect.

The goal of our study is to carry out accurate line profile calculations for
the \li\ doublet in strong magnetic fields. Two approaches were used to
evaluate the separation and relative strengths of the $\pi$ and $\sigma$
components of the Li line formed in the Paschen-Back regime. First, we have solved the
system of equations for the \mt\ multiplet following the formalism developed by
Darwin (\cite{D27}). Then, we have performed equivalent calculations with the
help of a computer program provided by E.~Landi Degl'Innocenti, who implemented
more advanced mathematical methods of the Racah algebra described in Landi
Degl'Innocenti \& Landolfi (\cite{LL04}). Results of the two alternative
computations are in excellent agreement. In Fig.~\ref{fig1} we illustrate the
separation and strength of the \mt\ multiplet components as a function of
$\omega/(\omega+1)$. In this scaled format the figure characterizes behaviour of
all astrophysically important \mt\ transitions (the \li\ resonance doublet, 
Na~D, Ca H and K, \ion{Mg}{ii} resonance lines). The vertical bars in
Fig.~\ref{fig1} indicate the magnetic field strength corresponding to a given 
splitting pattern for the Na~D lines and the \li\ doublet. It is clear that 
deviations from the linear Zeeman splitting for the latter feature can no longer be
neglected for $B$\,$\ga$\,2~kG, while significant asymmetry in the intensity of the
red and blue $\pi$ components appears already at $B$\,$\ga$\,1~kG.

We have incorporated results of the Paschen-Back effect modelling in the
polarized radiative transfer calculations of the \li\ 6708~\AA\ line profile.
The total oscillator strength of the Li multiplet is adopted from Yan et al.
(\cite{YTD98}). The presence of a non-zero nuclear spin implies that the lithium line
exhibits hyperfine splitting. This effect should be treated simultaneously with
the magnetic splitting only for the magnetic fields below 1~kG  (Traub \cite{T68}). In this
study we investigate the Li line in stars with much stronger fields, therefore a
self-consistent treatment of the hyperfine and magnetic splitting is
unnecessary. In fact, the separation of the hyperfine components is so small
that it can be neglected for all calculations presented here. The wavelengths
and relative strengths of the $^6$Li and $^7$Li line components are taken from
Smith et al. (\cite{SLN98}). Their line list is consistent with the very accurate
theoretical calculations (Sansonetti et al. \cite{SRE95}) and experimental
results (Volz \& Schmoranzera \cite{VS96}).

\begin{figure*}[!th]
\figps{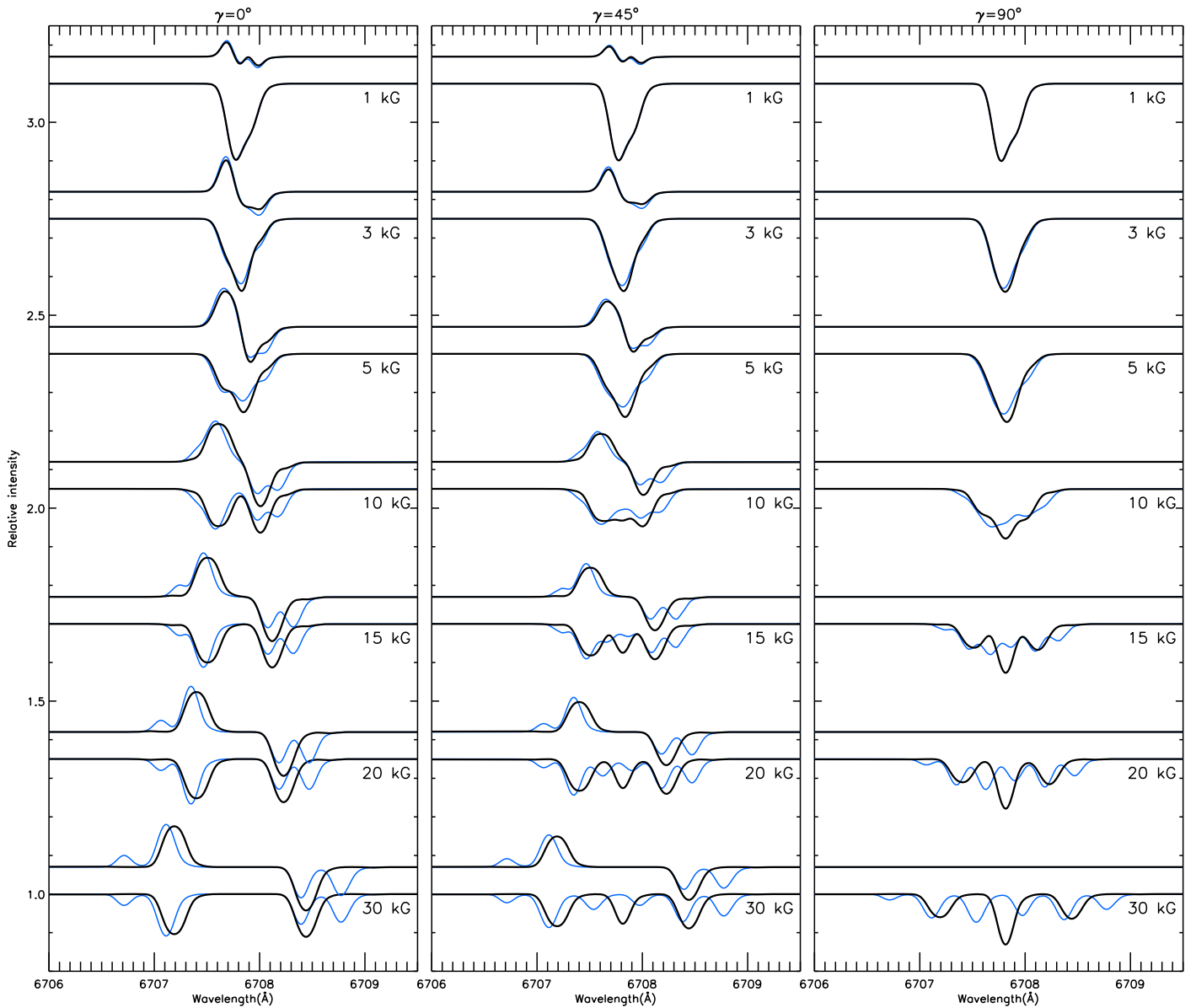}
\caption{Theoretical disk-centre Stokes $I$ and $V$ profiles of the \li\ 6708~\AA\
doublet computed with the \sm\ code. The three panels show the spectra for different
angles $\gamma$ between the magnetic field vector and the line of sight.
Calculations for different field strength ($B$\,=\,1, 3, 5, 10, 15, 20 and 30 kG) are shifted 
upward for display purposes.
The Stokes $V$ profiles are also shifted upward by 1.07 relative to the corresponding
Stokes $I$ spectra. The two sets of calculations are
shown: synthetic spectra including the Paschen-Back effect (thick curves) and the
calculations treating the Li line splitting in the Zeeman regime (thin curves).}
\label{fig2}
\end{figure*}

\begin{figure}[!th]
\figps{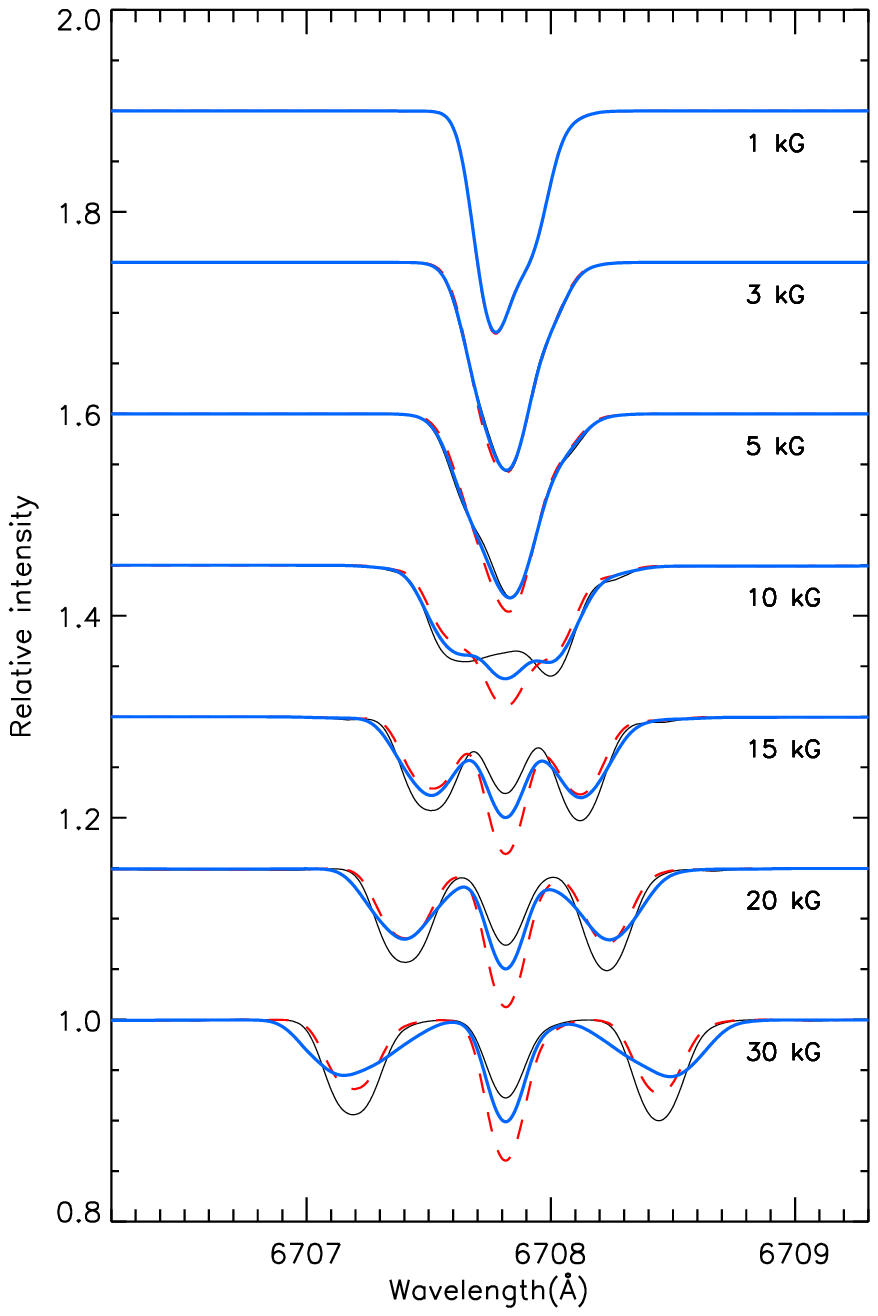}
\caption{Theoretical disk-integrated Stokes $I$ profiles of the \li\ 6708~\AA\
doublet. Calculations for different field strength (\bs\,=\,1, 3, 5, 10, 15, 20 and 30 kG) 
are shifted upward for display purposes. The
spectra computed for different magnetic field geometries are shown with the thin solid
line (homogeneous radial field), dashed line (homogeneous azimuthal field)
and the thick solid line (centred dipole field aligned with the line of sight).}
\label{fig2b}
\end{figure}

\subsection{Local profiles}
\label{local}

To avoid complications due to integration of the spectra over unresolved stellar
surface with a complex magnetic geometry, we first study the \li\ line profile for a
solar-like case. The local disk-centre theoretical Stokes $I$ and $V$ spectra of the
Li doublet are shown in Fig.~\ref{fig2}. These calculations, based on the polarized
radiative transfer code \sm\ (Kochukhov \cite{synthmag}), employ the {\tt ATLAS9}
model atmosphere (Kurucz \cite{K93}) with \teff\,=\,8000~K, \lgg\,=\,4.0 and Li
abundance $\log (N_{\rm Li}/N_{\rm tot})=-8.00$. This 3~dex enhancement of the Li
concentration relative to the solar photospheric Li abundance ($\log (N_{\rm
Li}/N_{\rm tot})=-10.99$, Asplund et al. \cite{as05}) is representative of the
abundance of this element inferred for the Li-rich Ap stars. The $^6$Li/$^7$Li
isotopic ratio is fixed at the solar system value of 0.07 (Anders \& Grevesse
\cite{AG89}). For the purpose of comparison with the standard treatment of the \li\ line
used in all previous studies of this feature in Ap stars, Fig.~\ref{fig2} also shows
line profile calculations neglecting the Paschen-Back effect. Polarized line profile
synthesis shows that in the magnetic fields stronger than $\approx$\,3~kG an accurate
modelling of the \li\ doublet is impossible without considering the Paschen-Back
effect. For the magnetic fields stronger than about 10~kG the line profile shape
predicted by the Zeeman regime calculations is entirely different from the correct
Paschen-Back treatment. 

\subsection{Disk-integrated profiles}
\label{disk}

A large variety of the global magnetic field geometries can be considered for Ap stars.
An assumption of the low-order multipolar topology is often used to fit the curves of the
magnetic observables (e.g. Landstreet \& Mathys \cite{LM00}). However, such models
cannot reproduce high-resolution observations of Ap stars in all four Stokes parameters
(Bagnulo et al. \cite{BWD01}; Kochukhov et al. \cite{KBW04}) and typically fail to achieve a
good fit to the Zeeman split lines 
(Bagnulo et al. \cite{BLL03}). Furthermore, for the slowly rotating cool Ap stars considered here no
phase-resolved high-resolution observations in the Stokes $I$ or other Stokes parameters are available.
In this situation multipolar models are severely underconstrained.

An alternative to using purely theoretical and often inadequate multipolar geometries
is to minimize the number of free parameters describing the magnetic field structure
and consider only those quantities which can be \textit{directly constrained} by the
available observational data. From the splitting of the Zeeman resolved lines one can 
determine the mean magnetic field strength, while the relative intensities of the $\pi$
and $\sigma$ components contain information about the field orientation. Thus, for the
purpose of calculation of the disk-integrated Stokes $I$ spectra it is convenient to
represent the field structure with a homogeneous distribution over the stellar surface.
This simple magnetic topology is characterized by a single value of the field strength
and a single inclination with respect to the stellar surface. Here we parameterize the field
with a combination of the radial and azimuthal field components. Note that the latter is
tangential to the stellar surface and is always perpendicular to the observer's line of sight.
The line of sight
symmetry of such homogeneous magnetic field geometry allows one to obtain disk-integrated
intensity spectra by combining the local \sm\ profiles, evaluated at 7 positions
between the disk-centre and the stellar limb (Kochukhov \cite{synthmag}). This
simplified treatment of the magnetic field topology of slowly rotating Ap stars is
usually adequate for the purpose of abundance analysis and the line profile fitting
(e.g. Kochukhov et al. \cite{KLR02}; Nielsen \& Wahlgren \cite{NW02}; 
Kochukhov \cite{K03}; Leone et al. \cite{LVS03}; Shavrina et al.
\cite{shav2}; Ryabchikova et al. \cite{RNW04,RRKB06}). In fact, a recent study
(Kochukhov \cite{K07b}) of the triplet lines in the spectra of strongly magnetic Ap
stars revealed that the narrow $\sigma$ components in a few such objects indicate a much
smaller surface scatter of the field strength compared to the predictions of low-order
multipolar magnetic models. This provides an additional justification for the simplified
single field strength value approach to synthesizing the Stokes $I$ spectra of magnetic
A stars.

Eventually, the homogeneous field assumption should fail when a large range of the
magnetic field strength variation over the visible stellar surface is present, inducing a broadening
of the $\sigma$ components. The only known observational example of this situation  in a
cool Ap star is found for HD\,154708 -- a star with an extremely strong magnetic field
(Hubrig et al. \cite{hubrig}).
Kochukhov (\cite{synthmag}) has shown that a major improvement in the fit to the triplet profile of the
\ion{Y}{ii} 5087~\AA\ line in this star can be achieved with a dipolar field aligned with
the line of sight. The Stokes $I$ spectra corresponding to this magnetic geometry can
be modelled with the help of a simple modification of the disk integration procedure
applied to the local profiles produced with \sm. Instead of a constant field strength and
field inclination values in 7 concentric rings, we can adopt parameters appropriate for the
aligned dipole.

In this section we present calculation of the \li\ line intensity for 7 values
of the mean field modulus, \bs, ranging from 1 to 30~kG. The model atmosphere, atomic parameters
and abundances are identical to those adopted previously. Calculations are carried out
for the extreme cases of the homogeneous field structure (radial field and azimuthal
field) and for the dipolar geometry aligned with the line of sight. In the latter case
the polar field strength of the dipole is adjusted to yield a given value of \bs. 
The stellar rotation and the instrumental broadening are neglected. 

The resulting profiles of the Li line are
illustrated in Fig.~\ref{fig2b} (only magnetic synthesis accounting for the partial
Paschen-Back effect is shown). It is evident that the morphology of the lithium line
profile is not very different from the local case considered in Sect.~\ref{local},
although variation of the field inclination with respect to the line of sight and the
surface changes of the field strength for the dipolar geometry introduce some smoothing of the
calculated spectra.

One can notice that the surface variation of the magnetic field strength in the dipolar
model also leads to additional differential broadening of the $\sigma$ components when
those become resolved. But for all the three field structures considered, the
characteristic simple triplet splitting pattern emerges in the 6708~\AA\ line for the
mean field modulus above $\approx$\,12~kG. This behaviour can be a key factor to support or refute
the Li line identification, provided that an absorption in the 6708~\AA\ region can be
detected for an Ap star with a strong enough magnetic field. 

\begin{figure}[!th]
\fifps{8.5cm}{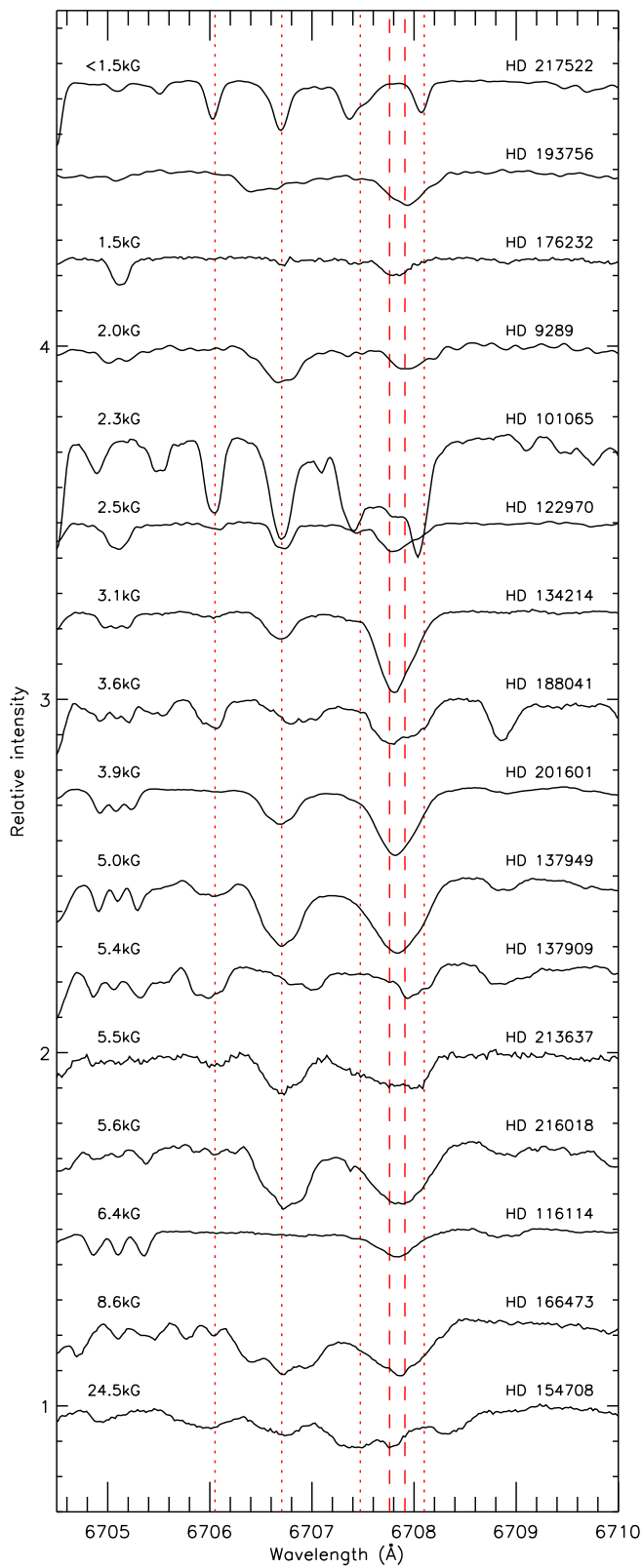}
\caption{Spectra of magnetic Ap stars in the 6708~\AA\ region. Stars are
arranged from top to bottom in the order of increasing magnetic field strength. 
The vertical lines 
show position of the \li\ doublet (dashed lines), as well as positions of some
prominent rare-earth lines, \ion{Ce}{ii} 6706.051, \ion{Pr}{iii} 6706.703, 
\ion{Sm}{ii} 6707.473, and \ion{Ce}{ii} 6708.099~\AA\ (dotted lines).}
\label{fig3}
\end{figure}

\section{The Li line in magnetic Ap stars}
\label{lihd}
 
To clarify the question of the presence of Li in magnetic Ap stars and to choose the
best objects for application of the theoretical framework outlined in Sect.~\ref{pbe},
we examined the high-resolution spectra of 17 slowly rotating late A magnetic peculiar
stars. All observations analysed in the present study were obtained with the UVES
spectrograph at the ESO VLT, using either the dichroic setup (390+580~nm) with a
0.5$^{\prime\prime}$ slit or the 600~nm red arm setting with an image slicer. The
resolving power reaches 80\,000 for the slit spectra and 115\,000 for the observations
obtained with the image slicer, respectively. Detailed description of the reduction
procedures employed for both types of the UVES data can be found in Kochukhov et al.
(\cite{KTR06}, \cite{KRW07}). All observations in the 600~nm red-arm mode were
acquired in the context of the programs devoted to the time-resolved monitoring of
pulsations in rapidly oscillating Ap (roAp) stars. These spectral time series, containing typically 60--150
individual exposures, were combined to yield $S/N$\,$\ge$\,300  for most of the
targets. Comparable noise levels were also achieved for the UVES slit spectra.
One-dimensional extracted and merged spectra were rectified in the 6680--6730 and
6130--6170~\AA\ regions using low-order polynomials. The first spectral interval is
employed in the analysis of the \li\ doublet; the second segment is useful for the determination
of the average  magnetic field strength from the separation of the resolved Zeeman 
components of the magnetically-sensitive \ion{Fe}{ii} line at $\lambda$ 6149~\AA\
(Mathys et al. \cite{mathys97}).

The sample of 17 objects studied here includes 9 sharp-lined Ap stars in which
the presence of the 6708~\AA\ feature was previously reported in the literature
(Faraggiana et al. \cite{FGD96}; Polosukhina et al. \cite{polo}; Kochukhov \cite{K03}). 
The three broad-lined stars
with a strong variable 6708~\AA\ line (HD\,3980, HD\,60435 and HD\,83368) are not
considered in our study because their relatively large projected rotational
velocities and the prominent horizontal inhomogeneities prohibit an accurate investigation
of the subtle magnetic broadening and splitting effects.

A collection of the 16 spectra of cool Ap stars in the Li region is presented in
Fig.~\ref{fig3}. The stars are arranged according to the mean magnetic field strength,
which is taken from the literature or determined using the  \ion{Fe}{ii} 6149~\AA\
line. The spectra are shifted to the laboratory frame using 3--5 lines of the iron-peak
and rare-earth elements in the 6690--6720~\AA\ region. All stars except HD\,217522
clearly show an absorption at the expected position of the \li\ 6708~\AA\ line. The
presence of this feature in the spectra of HD\,9289, HD\,116114, HD\,122970,
HD\,154708, HD\,193756 and HD\,216018 is reported here for the first time.

From the comparison of the 6708~\AA\ line with the strength of the nearby lines of
\ion{Nd}{iii}, \ion{Pr}{iii}, \ion{Ce}{ii}, \ion{Gd}{ii} and \ion{Sm}{ii} one can
conclude that the former line does not exhibit an obvious correlation with the
absorption due to any known REE features, thus making an alternative REE
identification of the 6708~\AA\ line unlikely. The cases of HD\,217522 and HD\,75445,
in which no Li line is detected, provide a particularly convincing illustration of the
deviating behaviour of the 6708~\AA\ line. The former star is one of the coolest
magnetic Ap stars known. It shows strong lines of the REEs in several ionization stages.
All four rare-earth lines in the vicinity of the \li\ feature are prominent in the
spectrum of HD\,217522, yet no absorption at the position of the lithium resonance
doublet is observed. 

\begin{figure}[!th]
\figps{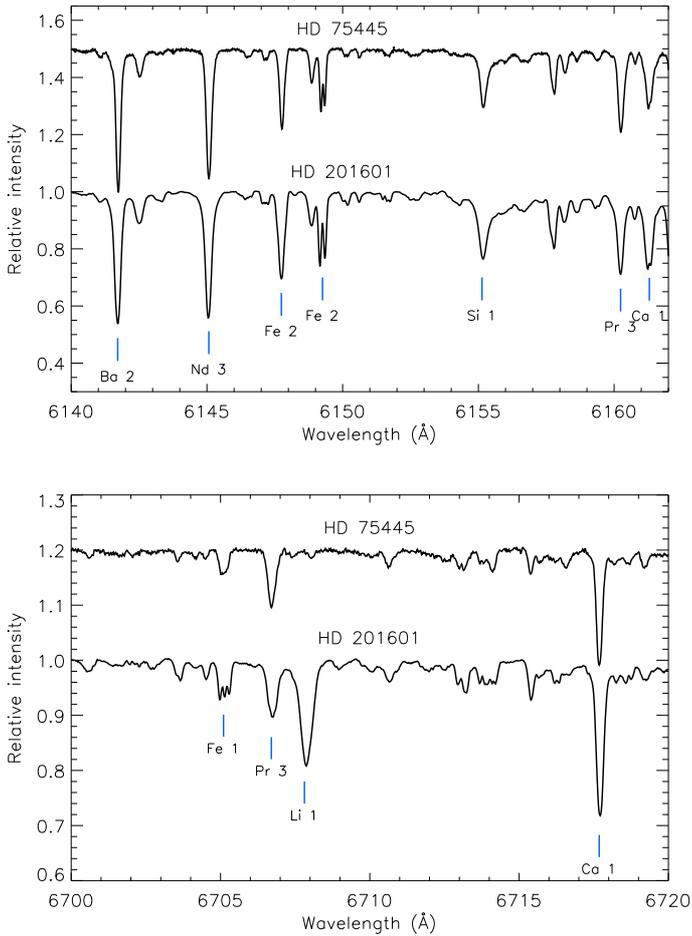}
\caption{Comparison of the spectra of HD\,75445 and HD\,201601 in the
6140--6162~\AA\ (upper panel) and 6700--6720~\AA\ (lower panel) regions. The 
two stars have very similar abundances and the line absorption, except for the
6708~\AA\ line, which is strong in HD\,201601 but absent in HD\,75445.}
\label{fig4}
\end{figure}

The Ap star HD\,75445 appears to be very similar to the bright roAp star HD\,201601
($\gamma$~Equ). Comparison of the stellar spectra in the 6140--6162~\AA\ region
(Fig.~\ref{fig4}, upper panel) shows little difference apart from the marginal effect
caused by a slightly stronger magnetic field in $\gamma$~Equ. The abundance analyses
by Ryabchikova et al. (\cite{RPK02,RNW04}) show that the two stars closely
resemble each other in the atmospheric parameters and chemical composition.
Nevertheless, the two stars exhibit remarkable discrepancy in the \li\ region:
$\gamma$~Equ shows a strong line at $\lambda$~6708~\AA, while this line is absent
entirely in HD\,75445 (Fig.~\ref{fig4}, lower panel). This finding again supports the
view that the 6708~\AA\ feature cannot be produced by an unidentified line of a common
REE ion.

The sequence of spectra presented in Fig.~\ref{fig3} highlights the difficulty of
quantitative analysis of the Li region. The Li line is often blended by \ion{Sm}{ii}
6707.473 and \ion{Ce}{ii} 6708.099~\AA, most prominently in HD\,101065. In stars with
stronger fields, the wings of the \li\ line and the resolved Zeeman components of
\ion{Pr}{iii} 6706.703~\AA\ start to overlap, further complicating the analysis.

It is not the intention of this paper to perform detailed modelling of the
6708~\AA\ region and to present accurate Li abundance determination for every
star in Fig.~\ref{fig3}. Instead we wish to answer the question if theoretically
predicted Paschen-Back splitting signature of the Li line can be observed, thus
providing a strong argument to support the identification of the 6708~\AA\
feature with the resonance \ion{Li}{i} line. Previous studies of the Li region 
in Ap stars (Shavrina
et al. \cite{shav1,shav2}; Polosukhina \& Shavrina \cite{PS07}) have
succeeded in obtaining reasonable fits to the Li blend neglecting the Paschen-Back
effect for the stars with fields $\le$\,6~kG. This was achieved by introducing a number
of additional fitting parameters (different types of broadening, field strength,
field inclination
were adjusted on the line-by-line basis, additional low-quality predicted
REE  lines were used). This means that, although the Paschen-Back effect could be
important for these stars, its observational signature is ambiguous. Therefore, in the
context of our investigation, we will study in detail only a few strong-field stars, 
showing a particularly distinct Li line profile shape or exhibiting 
resolved components of the Li blend.

Magnetic splitting of the resonance Li line is not observed in any star except,
possibly, in HD\,154708 (\bs\,=\,24.5~kG). This is in a qualitative agreement with the
theoretical predictions of the behaviour of the Li line in magnetic field. In
Sect.~\ref{pbe} we found that even in the most favourable cases of negligible
rotation, resolved components of the \li\ doublet should appear only for the fields
exceeding $\approx$\,12~kG. The two other stars, distinguished by the
strong magnetic fields and by the appearance of their spectra in the Li region, are
HD\,116114 and HD\,166473. These objects have the third and the second strongest fields,
respectively, among the
Ap stars analysed here and, at the same time, show a moderately strong \li\ feature.
Furthermore, in HD\,116114 this line
is free from interfering absorption of the REEs. Thus, we choose HD\,116114, HD\,166473 and
HD\,154708 for the detailed spectrum synthesis analysis based on the Paschen-Back
calculations of the magnetic splitting in the \li\ 6708~\AA\ line.

Our magnetic spectrum synthesis is carried out for the
two types of the magnetic field geometries discussed above
(homogeneous and pole-on dipolar). The relevant magnetic
field parameters and \vs\  are adjusted with the help of lines with a simple
and distinct Zeeman patterns (triplets). We do not attempt to tune these
parameters further in order to improve the Li line description. Instead, the observed
spectra are fitted by varying only the element abundances and macroturbulent broadening.
The latter is required to account for the broad line profiles of some ions in
cool Ap stars (Kochukhov \cite{K03}; Ryabchikova et al. \cite{RSK07}). 
The reality of turbulent broadening is also
supported by the analysis of line profile variations in roAp stars
(Kochukhov et al. \cite{KRW07}).
On the other hand, the lack of diagnostic lines and the absence of time-resolved spectra
does not allow us to consider chemical
stratification and spotted distribution of Li.  For these
reasons, and because we adopt a simplified description of the magnetic field
structure, our theoretical calculations will not give an ideal fit to
observations. However, we believe that the basic qualitative  agreement between
the observed and synthesized spectra is sufficient to verify the behaviour of
the 6708~\AA\ line in strong magnetic fields.

\subsection{HD\,116114}

The cool Ap star \hd\ shows resolved Zeeman split lines, indicating a mean
field modulus, \bs, of 5.9--6.0~kG and a long rotation period (Mathys et al.
\cite{mathys97}). In a later investigation Landstreet \& Mathys (\cite{LM00})
adopted a 27.6~d rotation period and fitted the phase curves of magnetic
observables with a global dipolar-like field, forming a small angle with the
rotational axis of the star. At the same time, the rotational modulation of all
magnetic observables is marginal and could still be consistent with a much
longer rotation period.

The atmospheric properties of \hd\ were investigated by Ryabchikova et al.
(\cite{RNW04}). They determined \teff\,=\,8000~K, \lgg\,=4.1, \bs\,=6.2~kG and
found a moderately large overabundance of the iron-peak elements for this star.
Compared to many other Ap stars in this temperature range, \hd\ does not
exhibit enhanced absorption due to the \ion{Nd}{iii} and \ion{Pr}{iii} features.
Consequently, the contribution of the \ion{Pr}{iii} 6706.703~\AA\ line in the Li
region is unimportant, which facilitates accurate analysis of the \li\ line. 
For the field strength of \hd\ we expect to see no splitting of the \li\ line
(see Figs.~\ref{fig2} and \ref{fig2b}), but the partial Paschen-Back effect is already quite
pronounced: the Li blend is deeper, narrower and more symmetric, 
the red wing is less extended, and the central structure is absent.

\begin{figure}[!th]
\figps{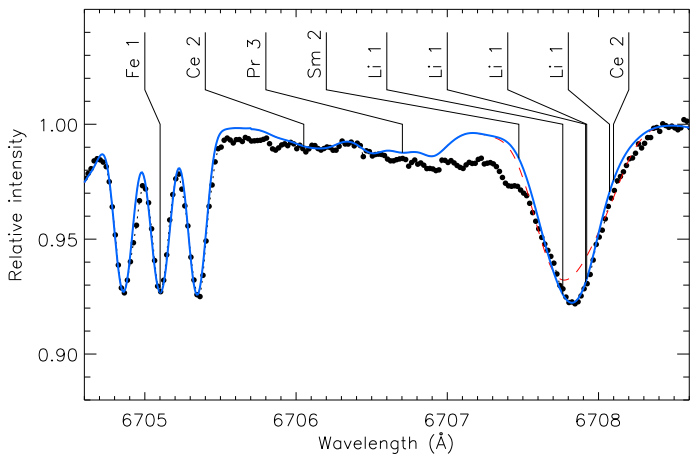}
\caption{Comparison of the observed (symbols) and computed (curves) spectra of
the cool Ap star \hd. The solid curve shows calculations accounting for the
partial Paschen-Back effect in the \li\ 6708~\AA\ line, while the dashed curve 
illustrates the synthetic spectrum computed for the Zeeman treatment of the 
\li\ line splitting.}
\label{fig5}
\end{figure}

We have performed a spectrum synthesis modelling of the Li region of \hd\ with the 
help of the \sm\ code, and using the element abundances and model atmosphere 
of Ryabchikova et al.
(\cite{RNW04}). The resonance lithium doublet is treated in the same way as in
Sect.~\ref{pbe}. The atomic parameters of other lines in the Li region are extracted
from the VALD database (Kupka et al.~\cite{VALD2}), which includes the DREAM data
for rare-earth elements (Bi\'emont et al. \cite{dream}).
The resulting line list is used throughout the present paper.
The average field strength and the field  orientation in \hd\
were determined by fitting the resolved Zeeman components of the \ion{Fe}{i}
6705.101~\AA\ line. A good description of the observed profile of this feature in
\hd\ (Fig.~\ref{fig5}) is achieved for \bs\,=\,6.4~kG, a 40\degr\ inclination of
the field vector relative to the surface normal and the projected rotational velocity
\vs\,=\,3.5~\kms. Theoretical calculations for the \li\ line using these magnetic
field parameters and \vs\ are presented in Fig.~\ref{fig5}. A good agreement of
observations and spectrum synthesis is found for  $\log (N_{\rm Li}/N_{\rm
tot})=-8.35$ and a contribution of the light lithium isotope similar to the solar
system value $^6$Li/$^7$Li=0.07 (Anders \& Grevesse \cite{AG89}). A macroturbulent
broadening of $V_{\rm mac}$\,=\,7~\kms\ is applied to the \li\ line, while the
\ion{Fe}{i} 6705.101~\AA\ line indicates $V_{\rm mac}$\,$\la$\,2~\kms. This range
of $V_{\rm mac}$ inferred from the lines of different elements agrees with the 
results obtained by Ryabchikova et al. (\cite{RSK07}).

A comparison of
the spectral synthesis for the Zeeman and partial Paschen-Back treatment of the
\li\ line splitting shows that the latter approach 
is more successful in
reproducing the 6708~\AA\ feature in the spectrum of \hd. In particular, the
Paschen-Back calculations predict a narrow line core, which agrees better with the
observed profile shape. 
At the same time, the Paschen-Back calculations lead to a 
deterioration of the fit in the red wing. This problem may be
related to an unaccounted blending or to a spotted distribution of Li. Despite this
uncertainty, it can be concluded that observations of \hd\ are
consistent with the \li\ identification and confirm theoretically expected
partial Paschen-Back behaviour of the magnetic splitting in the lithium resonance
doublet.

\subsection{HD\,166473}

The cool Ap star HD\,166473 shows an unusually large range of the field modulus variation with
a period of about 10~yr (Mathys et al. \cite{mathys97}). Our observations of this star were
obtained close to the maximum of \bs\ and, thus, are best-suited for the investigation of the
Paschen-Back effect in the \li\ line. The lithium feature in HD\,166473 was modelled
previously by Polosukhina \& Shavrina (\cite{PS07}).
They have neglected the Paschen-Back effect and used observations corresponding to
the rotational phase of weaker \bs.

\begin{figure}[!th]
\figps{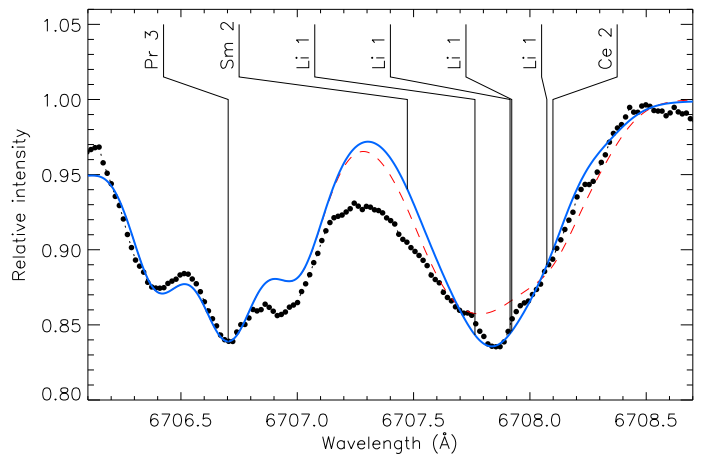}
\caption{The same as Fig.~\ref{fig5} for HD\,166473.}
\label{fig6}
\end{figure}

Gelbmann et al. (\cite{GRW00}) analysed in detail the atmospheric properties of HD\,166473.
They have determined the stellar parameters, \teff\,=\,7700~K and \lgg\,=\,4.2, and
have inferred a large overabundance of the REEs along with a moderate enhancement of the
iron-peak elements. The effective temperature obtained by Gelbmann et al. (\cite{GRW00}) was
subsequently confirmed on the basis of detailed analysis of the hydrogen Balmer line profiles
(Kochukhov et al. \cite{KBP02}).

Our theoretical calculations of the \li\ line (Sect.~\ref{pbe}) show that for the field strength
of HD\,166473 and an insignificant line broadening one should expect to find a large influence of the
partial Paschen-Back effect. In particular, for the fields inclined significantly with
respect to the line of sight, the Li line profile acquires a distinct triangular shape,
similar to the one observed in the stellar spectrum.

Here we have modelled the \li\ resonance doublet in HD\,166473 using the model atmosphere
parameters and abundances from Gelbmann et al. (\cite{GRW00}). The model atmosphere of
HD\,166473 was calculated with the {\tt LLModels} code (Shulyak et al. \cite{llmodels}),
taking into account individual stellar chemical composition and including the approximate
treatment of the Zeeman effect in the line opacity (Kochukhov et al. \cite{KKS05}).

The comparison of observations and the spectrum synthesis of the \ion{Pr}{iii} 6706.703~\AA\
+ \li\ blend (Fig.~\ref{fig6}) suggests that the mean magnetic field vector should be inclined
by $\approx$\,70\degr\ with respect to the stellar surface normal for the field \bs\,=\,8.6~kG
to match the deep central component of the \ion{Pr}{iii} line. At the same time, a large
macroturbulent broadening, $V_{\rm mac}$\,=\,9~\kms, is needed to fit the line width. Due to
the long rotation period of HD\,166473, the rotational Doppler broadening can be neglected.
The fit illustrated in Fig.~\ref{fig6} was produced using 
$\log(N_{\rm Pr}/N_{\rm tot})=-7.20$, $\log(N_{\rm Li}/N_{\rm tot})=-8.05$ and adopting
Gelbmann et al. (\cite{GRW00}) abundances for other elements. The oscillator strength of the
\ion{Pr}{iii} line, $\log gf=-1.64$, is from Bi\'emont et al. (\cite{dream}).
The inferred \ion{Pr}{iii} abundance is consistent with Gelbmann et al. (\cite{GRW00})
estimate of $\log(N_{\rm \ion{Pr}{iii}}/N_{\rm tot})=-7.60\pm0.38$.

Theoretical calculations including the Paschen-Back effect lead to a somewhat better
description of the observations compared to the synthesis based on the linear Zeeman splitting
of the Li blend components. The latter calculation is unable to match the narrow triangular
core of the \li\ line even for highly inclined fields. However, significant missing opacity in
the 6706.9--6707.5~\AA\ interval and the large broadening complicates the assessment of the
partial Paschen-Back effect signature in HD\,166473.

\subsection{HD\,154708}

A very strong magnetic field, \bs\,=\,$24.5\pm1.0$~kG, was detected in Ap
star HD\,154708 by Hubrig et al. (\cite{hubrig}). Their \teff\
determination led to contradictory results, with the $B2-G$ temperature
calibration of Geneva photometry (Hauck \& North \cite{HN93}) giving
\teff\,=\,6800~K, while Str\"omgren photometry (Napiwotzki et al. \cite{N93}) 
pointing to \teff\,=\,7500~K. We find that the hydrogen Balmer lines in
HD\,154708 are better reproduced with the latter \teff\ value and that Geneva
photometric indices are known to be susceptible to anomalous line blanketing
(Kochukhov et al. \cite{KKS05}). Hence, here we adopt \teff\,=\,7500~K for the
spectrum synthesis analysis of HD\,154708. 
The model atmosphere for this very peculiar star is
calculated with the {\tt LLModels} code, assuming \lgg\,=\,4.0. This model
takes into account 
preliminary abundance estimates derived from fitting metal
lines in the two studied spectral regions and includes the effect of magnetic
field on the metal line opacity.

A comparison of the theoretical spectra for a single value of magnetic
field strength with the observed profiles of triplet lines shows that a
homogeneous magnetic topology, similar to the one used for \hd\ and HD\,166473, is inadequate
for HD\,154708. While the $\pi$ components in the spectrum of this star are
rather narrow and can be fitted with \vs\,$\approx$\,5~\kms, the $\sigma$
components show an excessive broadening, indicating substantial spread of the
field modulus over the visible part of the stellar surface. 
To account for this
property of the field structure in HD\,154708, we have modified the disk
integration procedure applied to the local profiles produced with \sm\ in such
a way that the resulting flux spectra would correspond to a dipolar field aligned
with the line of sight (see Sect.~\ref{disk}). 
With this modification, a major improvement in the fit
to triplet lines can be achieved for a dipolar field structure with 
$B_{\rm p}$\,=\,30~kG
(see Kochukhov \cite{synthmag}). For this geometry the disk-averaged field strength is
\bs\,=\,24~kG, which agrees with the direct measurements.

\begin{figure}[!t]
\figps{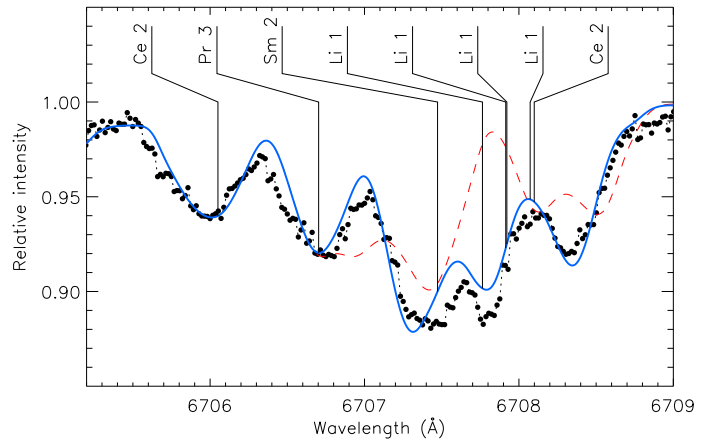}
\caption{The same as Fig.~\ref{fig5} for HD\,154708.}
\label{fig7}
\end{figure}

Inspection of the Li region shows that the feature which could possibly be
identified with the splitted \li\ line is blended by the strong line of
\ion{Pr}{iii} at $\lambda$ 6706.703~\AA\ (Fig.~\ref{fig7}). In the
6705.5--6709.0~\AA\ interval the spectrum of HD\,154708 shows 5 line components
with the central wavelengths 6706.0, 6706.7, 6707.4, 6707.8, and 6708.4~\AA. The
first two features can be attributed to the groups of the blueshifted $\sigma$ and $\pi$
components of the \ion{Pr}{iii} line. The other components contain significant
contribution of the absorption due to the \li\ line, which exhibits nearly
complete Paschen-Back splitting for the field strength of HD\,154708. Adopting the
magnetic field geometry described above, we are able to identify the feature
at $\lambda$ 6707.4 with the blend of the redshifted \ion{Pr}{iii} $\sigma$
components and the blueshifted $\sigma$ component of the \li\ triplet. The 
remaining two features belong to the $\pi$ and to the redshifted $\sigma$ components
of the Li line.

The spectrum synthesis results presented in Fig.~\ref{fig7} demonstrate a
reasonable agreement of the observed profile of the \ion{Pr}{iii} and \li\
blend with the theoretical calculations including the Paschen-Back effect for
\li\ 6708~\AA. These spectra were obtained adopting the solar system ratio of
the Li isotopes and using an enhanced total abundance of this element, $\log
(N_{\rm Li}/N_{\rm tot})=-8.20$. 
The REE abundances, $\log(N_{\rm Pr}/N_{\rm tot})=-7.70$, $\log(N_{\rm
Ce}/N_{\rm tot})=-8.20$, $\log(N_{\rm Sm}/N_{\rm tot})=-8.50$, were used.
The relevant REE line parameters (Bi\'emont et al. \cite{dream}) were extracted from
VALD. The optimal description of the observed spectrum is achieved for $V_{\rm
mac}=7$~\kms.

As expected, computation for the Zeeman
splitting pattern of the lithium doublet produces unrealistic spectrum showing
no resemblance to the actual observations. However, even calculations with the
detailed Paschen-Back treatment of the \li\ line do not reproduce 
some details, in particular the narrow 6707.8~\AA\ component, in the
HD\,154708 spectrum.
We attribute this
problem to the simplistic magnetic field topology adopted in our polarized
radiative transfer calculations. In reality the field structure probably
deviates from a pure dipole, and spectra may be influenced by the horizontal
and vertical
chemical inhomogeneities. Additional constraints on the magnetic and
abundance distributions in HD\,154708 require spectroscopic phase-resolved
monitoring of this star. Nevertheless, even with the limited data available in
the present study, it is remarkable to find that the absorption features
observed in the 6708~\AA\ region of HD\,154708 coincide with the theoretically predicted positions
of the magnetically-split components of the \li\ doublet. The latter behaves
unusually in the strong magnetic field of this star, and it is very unlikely that such a
coincidence (both in the central wavelength and in the splitting pattern) could be due
to an unrecognized blending. Thus, observations of HD\,154708 lend strong
support to the suggestion that the line at 6708~\AA\ in the spectra of cool Ap
stars is due to lithium, which is strongly overabundant in the line-forming atmospheric
region.

\section{Conclusions} 
\label{conc}

In this study we addressed the question of the unusual behaviour of the \li\
resonance doublet in a strong magnetic field. Due to the small separation of the Li
doublet components, its magnetic splitting occurs in the partial Paschen-Back regime
for the field strengths typically found at the surfaces of magnetic stars. We
model in detail the Paschen-Back splitting in the Li transition and present the
first polarized radiative transfer spectrum synthesis calculations that
take into account the partial Paschen-Back effect for this line.
Comparing these calculations with the high-quality spectra of cool magnetic Ap
stars, we are able to shed a new light on the long-standing problem of the \li\
6708~\AA\ line identification in Ap stars.

The main findings of our investigation can be summarized as follows:
\begin{itemize}
\item[$\bullet$] The presence of significant departures of the Li line splitting pattern
from the one expected for the linear Zeeman effect in a moderately strong field 
is hereby confirmed. The local line profile calculations show that the Paschen-Back
effect becomes noticeable in the Stokes $V$ for $B$\,$\ga$\,1~kG and 
and in the Stokes $I$ for $B$\,$\ga$\,3~kG, while a
major modification
of the \li\ line profile shape is found for $B$\,$\ge$\,5~kG. 
Since the magnetic fields
of the Li-rich Ap stars are often stronger than this limit, these results question the
validity of the previous attempts (e.g., Polosukhina \& Shavrina \cite{PS07}) to 
model the \li\ line ignoring the Paschen-Back effect.
\item[$\bullet$] Calculations of the disk-integrated Stokes $I$ spectra
show that the triplet splitting pattern emerges in the \li\ line for 
\bs\,$\ga$\,12~kG, offering a unique opportunity to verify the 6708~\AA\ line
identification in Ap stars with a very strong magnetic field.
\item[$\bullet$] A survey of the available high-resolution UVES spectra of cool Ap stars
reveals the presence of significant 6708~\AA\ absorption in 15 stars, out of
which 6 objects are identified as potentially Li-rich stars for the first
time.
\item[$\bullet$] The intensity of the 6708~\AA\ line does not correlate with the overabundance
of the REEs, thus invalidating the hypothesis that this line is produced
by an unidentified feature of one of the common REE ions. The cases of HD\,75445 and
HD\,217522, which do not show the \li\ absorption but are very similar in their
REE abundances to known Li-rich Ap stars, emphasize the difficulty of attributing
the 6708~\AA\ line to anything but Li.
\item[$\bullet$] We undertook 
spectrum synthesis modelling of the \li\ line in Ap
stars HD\,116114, HD\,166473 and HD\,154708. 
The latter object has the strongest
magnetic field among all cool Ap stars and, thus, represents a unique natural
laboratory for the study of the Paschen-Back effect in the resonance lithium
line. We find that the synthetic spectral calculations which properly 
account for the
Paschen-Back splitting improve the fit to the \li\ line in the spectra of
HD\,116114 and HD\,166473. For HD\,154708, theoretically expected Paschen-Back triplet splitting of the \li\
line is detected in the observed spectrum. Given the difficulty of treating the
unusually strong 
magnetic field of this star and the lack of observational constraints for the 
field geometry, the agreement between
the predicted and observed positions and relative strengths of the resolved magnetic
components of the blend containing the \li\ line is deemed satisfactory. For all three stars
we need roughly 2.6--2.9~dex overabundance of Li relative
to the chemical composition of the solar photosphere to fit the observed line
profiles.
\item[$\bullet$] In summary, we confirm the \li\ identification of the 6708~\AA\
line in cool Ap stars. The puzzling diversity of the strength of the Li absorption
in stars with similar atmospheric parameters suggests that this element is
sensitive to the time-dependent diffusion effects (Alecian \cite{A98}), or,
possibly, is influenced by the chemical weather
recently discovered in the
non-magnetic chemically peculiar stars (Kochukhov et al. \cite{aand}).
Interesting and diverse behaviour of Li in cool Ap stars 
challenges modern atomic diffusion theories (e.g., Alecian \& Stift \cite{AS06}) 
to find a satisfactory
explanation of the accumulation of this element in the line-forming atmospheric layers.
\end{itemize}

\begin{acknowledgements}
This study benefited from stimulating discussions with P. Barklem, 
S. Bagnulo, T. Ryabchikova and M. Stift. I thank E. Landi Degl'Innocenti for providing 
a computer program to calculate partial Paschen-Back splitting of spectral lines. 
Comments by the anonymous referee contributed significantly to the improvement
of this paper.
\end{acknowledgements}

\noteaddname

In a publication, which appeared after submission of the present paper, Stift et al.
(\cite{SFL08}) presented a theoretical study of the partial Paschen-Back effect for a number
of Fe and Cr lines observed in Ap stars. They have also calculated the Paschen-Back 
splitting for the \li\ line, illustrating their results with a figure similar to Fig.~1 of 
the present paper.

\end{document}